\documentclass[lettersize,journal]{IEEEtran}
\usepackage{amsmath,amsfonts}
\usepackage{algorithmic}
\usepackage{array}
\usepackage[caption=false,font=normalsize,labelfont=sf,textfont=sf]{subfig}
\usepackage{textcomp}
\usepackage{stfloats}
\usepackage{url}
\usepackage{verbatim}
\usepackage{graphicx}

\usepackage{makecell}
\usepackage{pifont}% http://ctan.org/pkg/pifont

\usepackage{enumitem}

\usepackage{flafter}
\usepackage{flushend}
\usepackage{orcidlink}

\def\BibTeX{{\rm B\kern-.05em{\sc i\kern-.025em b}\kern-.08em
    T\kern-.1667em\lower.7ex\hbox{E}\kern-.125emX}}
\usepackage{balance}
\begin{document}
\title{Streaming Technologies and Serialization Protocols: Empirical Performance Analysis}

\author{Samuel Jackson \orcidlink{0000-0001-5301-5095},
Nathan Cummings \orcidlink{0000-0003-4359-6337}, and 
Saiful Khan \orcidlink{0000-0002-6796-5670}
\thanks{
Samuel Jackson and Nathan Cummings are with
Computing Division, Culham Centre for Fusion Energy, Culham Science Centre, Abingdon, OX14 3EB, Saiful Khan is with 
Scientific Computing Department, Science and Technology Facilities Council, Rutherford Appleton Laboratory, Didcot, OX11 0QX.
This work has been submitted to the IEEE for possible publication. Copyright may be transferred without notice, after which this version may no longer be accessible.
}}

% \address[1]{Computing Division, Culham Centre for Fusion Energy, Culham Science Centre, Abingdon, OX14 3EB}
% \address[2]{Scientific Computing Department, Science and Technology Facilities Council, Rutherford Appleton Laboratory, Didcot, OX11 0QX}

% \author{IEEE Publication Technology Department
% \thanks{Manuscript created October, 2020; This work was developed by the IEEE Publication Technology Department. This work is distributed under the \LaTeX \ Project Public License (LPPL) ( http://www.latex-project.org/ ) version 1.3. A copy of the LPPL, version 1.3, is included in the base \LaTeX \ documentation of all distributions of \LaTeX \ released 2003/12/01 or later. The opinions expressed here are entirely that of the author. No warranty is expressed or implied. User assumes all risk.}}

% \markboth{Journal of \LaTeX\ Class Files,~Vol.~18, No.~9, September~2020}%
% {How to Use the IEEEtran \LaTeX \ Templates}

\maketitle

\begin{abstract}
Efficient data streaming is essential for real-time data analytics, visualization, and machine learning model training, particularly when dealing with high-volume datasets. Various streaming technologies and serialization protocols have been developed to cater to different streaming requirements, each performing differently depending on specific tasks and datasets involved. This variety poses challenges in selecting the most appropriate combination, as encountered during the implementation of streaming system for the MAST fusion device data or SKA's radio astronomy data. To address this challenge, we conducted an empirical study on widely used data streaming technologies and serialization protocols. We also developed an extensible, open-source software framework to benchmark their efficiency across various performance metrics. Our study uncovers significant performance differences and trade-offs between these technologies, providing valuable insights that can guide the selection of optimal streaming and serialization solutions for modern data-intensive applications. Our goal is to equip the scientific community and industry professionals with the knowledge needed to enhance data streaming efficiency for improved data utilization and real-time analysis.   
\end{abstract}

\begin{IEEEkeywords}
Data streaming, messaging systems, serialization protocols, web services, performance evaluation, empirical study, and applications.
\end{IEEEkeywords}

\section{Introduction}
\label{sec:introduction}

\IEEEPARstart{W}{ith} the exponential increase in data generation from large scientific experiments and the rise of data-intensive machine learning algorithms in scientific computing~\cite{heyMachineLearningBig2020}, traditional methods of data transfer are becoming inadequate. This trend necessitates efficient data streaming methods that allow end-users to access subsets of data remotely. Additionally, the drive for FAIR~\cite{wilkinsonFAIRGuidingPrinciples2016, rocca-serraFAIRCookbookEssential2023}, and open data mandates require that such data be publicly accessible to end users via the internet.

\begin{table}
\caption{\textbf{List of abbreviations used in this study.}}
\label{table}
\setlength{\tabcolsep}{3pt}
\begin{tabular}{|l|l|}

    \hline
    \thead{Abbreviation} & \thead{Definition} \\
    \hline
    ADIOS & ADaptable Input Output System \\
    AMQP & Advanced Message Queuing Protocol \\
    API & Application Programming Interface \\
    BSON & Binary JSON \\
    CBOR & Concise Binary Object Representation \\
    CCFE & Culham Centre for Fusion Energy \\
    CoAP & Constrained Application Protocol \\
    EXI & Efficient XML Interchange \\
    FAIR & Findable, Accessible, Interoperable, and Reusable \\
    gRPC & Google RPC \\
    HTTP & Hypertext Transfer Protocol \\
    JSON & Javascript Object Notation \\
    MAST & Mega-Ampere Spherical Tokamak \\
    MAST-U & Mega-Ampere Spherical Tokamak Upgrade \\
    MQTT & MQ Telemetry Transport \\
    P2P & Peer-to-peer \\
    ProtoBuf & Protocol Buffers \\
    PSON & Protocol JSON \\
    REST & Representational State Transfer \\
    RPC & Remote Procedural Call \\
    SKA & Square Kilometer Array \\
    STOMP & Simple Text Orientated Messaging Protocol \\
    TCP & Transmission Control Protocol \\
    UBJSON & Universal Binary JSON \\
    UKAEA & UK Atomic Energy Authority \\
    WAN & Wide Area Network \\
    XHTML & eXtensible HyperText Markup Language \\
    XML & eXtensible Markup Language \\
    XMPP & Extensible Messaging and Presence Protocol \\
    YAML & YAML Ain’t Markup Language \\
    ZMQ & ZeroMQ \\
    
    \hline

\end{tabular}
\label{table:abbr}
\end{table} 

MAST \cite{sykesFirstPhysicsResults2001, SJackson:2024:SoftwareX} was a fusion reactor in operation at the UKAEA, CCFE from 1999 to 2013 and its upgraded successor is MAST-U \cite{harrisonOverviewNewMAST2019}, which began operation in 2020. These facilities generate gigabytes of data per experimental run, with possibly many runs per day. 
However, the lack of public access to the archive of data produced by the MAST experiment and other fusion devices around the world has limited collaborative opportunities with international and industry partners. The pressing need to facilitate real-time data analysis \cite{churchillFrameworkInternationalCollaboration2021} and leverage recent advancements in machine learning \cite{ anirudh_2022_2023, pavoneMachineLearningBayesian2023} further emphasizes the necessity for efficient data streaming technologies. These technologies must not only handle the sheer volume of data but also integrate seamlessly with analytical tools.

In this paper, we extend the work conducted in~\cite{khanIEEEAccess2023} for the SKA's radio astronomy data streaming and visualization. We explore an array of different streaming technologies available. We consider the combination of two major choices of technology when implementing a streaming service: (a) the choice of a streaming system, which performs the necessary communication between two endpoints, and (b) the choice of serialization used to convert the data into transmittable formats. Our contributions are as follows:

\begin{itemize}
    \item We provide a comprehensive review of 11 streaming technologies and 13 serialization methods, categorized by their underlying principles and operational frameworks.

    \item We introduce an extensible software framework designed to benchmark the efficiency of various combinations of streaming technology and serialization protocols, assessing them across 11 performance metrics.

    \item By testing over 143 combinations, we offer a detailed comparative analysis of their performance across eight different datasets. 

    \item Our findings not only highlight the performance differentials and trade-offs between these technologies, but we also discuss the limitations of this study and potential directions for further research.
\end{itemize}

Through this comprehensive study, we aim to equip the scientific community with deeper insights into choosing appropriate streaming technologies and serialization protocols that can meet the demands of modern data challenges. 

The rest of this work is structured as follows. Section \ref{sec:related-work} offers a concise review of related literature. Section \ref{sec:background} presents an overview of various serialization protocols and data streaming technologies examined in this study. Section \ref{sec:empirical-study-design} outlines our experimental methodology, including the performance metrics considered, implementation specifics of our benchmark framework and the selection of datasets used for evaluation. Section \ref{sec:results} analyzes experimental results covering all performance metrics and datasets. Finally, sections \ref{sec:discussion} and \ref{sec:conclusion} provides a discussion of the findings and recommendations of technology choices. 
\section{Related Work}
\label{sec:related-work}

Khan et al.~\cite{khanIEEEAccess2023} evaluated the performance of streaming data and web-based visualization for SKA's radio astronomy data. They also conducted a limited analysis on the serialization, deserialization, and transmission latency of two protocols - ProtoBuf and JSON. Our work builds on their research by covering a more extensive range of combinations.

Proos et al. \cite{proos_performance_2020} consider the performance of three different serialization formats (JSON, Flatbuffers, and ProtoBuf) and a mixture of three different messaging protocols (AMQP, MQTT, and CoAP). They evaluate the performance using real ``CurrentStatus'' messages from Scania vehicles as JSON data payload data. They monitor communication between a desktop computer and a Raspberry Pi. They consider numerous evaluation metrics such as latency, message size, and serialization/deserialization speed.

The authors of \cite{friesel_data_2021} compare 10 different serialization formats for use with different types of micro-controllers and evaluate the size of the payload from each method. They test performance with two types of messages 1) JSON payloads obtained from ``public mqtt.eclipse.org messages'' and 2) object classes from smartphone-centric studies  \cite{petersen_smart_2017, sumaray_comparison_2012}.
 
Fu and Zhang \cite{fu_fair_2021} presented a detailed review of different messaging systems. They evaluate each method in terms of throughput and latency when sending randomly generated text payloads. They evaluate each method only on the local device to avoid bias from any network specifics. Orthogonal to our work, they are focused on evaluating the scaling of each system over a number of producers, consumers, and message queues.

Churchill et al.~\cite{churchillFrameworkInternationalCollaboration2021} explored using ADIOS~\cite{godoyADIOSAdaptableInput2020} for transferring large amounts of Tokamak diagnostic data from the K-STAR facility in Korea to the NERSC and PPPL facilities in the USA for near-real-time data analysis.

We differentiate our study from these related works by evaluating 1) A wide variety of different streaming technologies, both message broker-based and RPC-based. 2) considering a large number of data serialization formats, including text, binary, and protocol-based formats. 3) We evaluate the combination of these technologies, developing an extensible framework for measuring and comparing serialization and streaming technologies. 4) Evaluating the performance over 11 different metrics. We comprehensively evaluate 11 different streaming technologies with 13 different serialization methods over 8 different datasets.

\section{Background}
\label{sec:background}

In this paper, we study how the choice of streaming technologies and serialization protocols critically affects data transfer speed. Specifically, we analyze the application of popular messaging technologies and serialization protocols across diverse datasets used in machine learning. Before discussing our experimental setup and results, this section provides an overview of message systems and serialization protocols suitable for streaming data. 

\subsection{Serialization Protocols}
\label{subsec:serialization-protos}
 
\begin{table*}[t]
\centering
\caption{A comparison of various serialization protocols. Type: describes how the method serializes data, e.g., text, binary, common protocol. Human readable: indicates whether the serialization scheme is legible to a human reader. Defined schema: specifies whether producer and consumer share a common knowledge of the data format prior to transmission. Code generated schema: states whether the serialization requires code to be generated from a predefined protocol.}

\setlength{\tabcolsep}{3pt}

\begin{tabular}{ |l|c|c|c|c|c|}
 
\hline

 \thead{Protocol} & \thead{Type} & \thead{Binary} & \thead{Human Readable} & \thead{Defined Schema}  & \thead{Code Generated Schema} \\

\hline 

XML \cite{noauthor_extensible_nodate} & Text & \ding{53} & \checkmark & \ding{53} & \ding{53} 
\\

JSON \cite{bray_javascript_2017} & Text & \ding{53} & \checkmark & \ding{53} & \ding{53} 
\\

YAML \cite{noauthor_yaml_nodate} & Text & \ding{53} & \checkmark & \ding{53} & \ding{53} \\

BSON \cite{noauthor_bson_nodate} & Binary & \checkmark & \ding{53} & \ding{53} & \ding{53} \\

UBSON \cite{noauthor_universal_nodate} & Binary & \checkmark & \ding{53} & \ding{53} & \ding{53} \\

CBOR \cite{noauthor_cbor_nodate} & Binary & \checkmark & \checkmark & \ding{53} & \ding{53}\\

MessagePack \cite{noauthor_messagepack_2023} & Binary  & \checkmark & \ding{53} & \ding{53} & \ding{53} \\

Pickle \cite{noauthor_pep_nodate} & Binary  & \checkmark & \ding{53} & \ding{53} & \ding{53} \\

ProtoBuf \cite{noauthor_protocol_nodate} & Protocol  & \checkmark & \ding{53} & \checkmark & \checkmark \\

Thrift \cite{slee_thrift_nodate} & Protocol  & \checkmark & \ding{53} & \checkmark & \checkmark \\

Capn'Proto (capnp, capn-packed)\cite{noauthor_capn_nodate} & Protocol  & \checkmark & \ding{53} & \checkmark & \checkmark \\

Avro \cite{noauthor_apache_nodate-1, sumaray_comparison_2012, friesel_data_2021} & Protocol &  \ding{53} & \checkmark & \checkmark & \ding{53} \\

\hline

\end{tabular}
\label{table:protocols}
\end{table*}

In this section, we provide a brief overview of three different categories of serialization protocol: text formats, binary formats, and protocol formats.

\subsubsection{Text Formats}
eXtensible Markup Language (XML)~\cite{noauthor_extensible_nodate} is a markup language and data format developed by the World Wide Web consortium. It is designed to store and transmit arbitrary data in a simple, human-readable format. XML adds context to data using tags with descriptive attributes for each data item. It has been extended to various derivative formats, such as XHTML and EXI.

JSON~\cite{bray_javascript_2017} is another human-readable data interchange format that represents data as a collection of nested key-value pairs. JSON is commonly used for data exchange protocol in RESTful APIs. Due to the smaller payload size, it is often seen as a lower overhead alternative to XML for data interchange.

YAML Ain’t Markup Language (YAML)~\cite{noauthor_yaml_nodate} is a simple text-based data format often used for configuration files. It is less verbose than XML and supports advanced features such as comments, extensible data types, and internal referencing. 

\subsubsection{Binary Formats}

Binary JSON (BSON)~\cite{noauthor_bson_nodate} is a binary data format based on JSON, developed by MongoDB. Similar to JSON, BSON also represents data structures using key-value pairs. It was initially designed for use with the MongoDB NoSQL database but can be used independently of the system. BSON extends the JSON format with several data types that are not present in JSON, such as a datetime format.

Universal Binary JSON (UBJSON)~\cite{noauthor_universal_nodate} is another binary extension to the JSON format created by Apache. UBJSON is designed according to the original philosophy of JSON and does not include additional data types, unlike BSON.

Concise Binary Object Representation (CBOR)~\cite{noauthor_cbor_nodate} is also based on the JSON format. The major defining feature of CBOR is its extensibility, allowing the user to define custom tags that add context to complex data beyond the built-in primitives.
 
MessagePack~\cite{noauthor_messagepack_2023} is a binary serialization format, again based on JSON. It was designed to achieve smaller payload sizes than BSON and supports over 50 programming languages.

Pickle \cite{noauthor_pep_nodate} is a binary serialization format built into the Python programming language. It was primarily designed to offer a data interchange format for communicating between different Python instances.

\subsubsection{Protocol Formats}

Protocol Buffers (ProtoBuf)~\cite{noauthor_protocol_nodate} were developed by Google as an efficient data interchange format, particularly optimized for inter-machine communication. Specifically, ProtoBuf is designed to facilitate remote procedural call (RPC) communication through gRPC~\cite{noauthor_grpc_nodate}. Data structures used for communication are defined in .proto files, which are then compiled into generated code for various supported languages. During transmission, these data structures are serialized into a compact binary format that omits names, data types, and other identifiers, making it non-self-descriptive. Upon receipt, the messages are decoded using the shared protocol buffer definitions.

Thrift~\cite{slee_thrift_nodate} is another binary data format developed by Apache Software Foundation or Apache, similar in many respects to ProtoBuf. In Thrift, data structures are also defined in a separate file, and these definitions are used to generate corresponding appropriate data structures in various supported languages. Before transmission, data are serialized into a binary format. Thrift is also designed for RPC communication and includes methods for defining services that use Thrift data structures. However, Thrift has a smaller number of supported data types compared to ProtoBuf. 

Capn'Proto~\cite{noauthor_capn_nodate} is a protocol-based binary format that competes with ProtoBuf and Thrift. Capn'Proto differentiates itself with two main features. First, its internal data representation is identical to its encoded representation, which eliminates the need for a serializing step. Second, its RPC service implementation offers a unique feature called ``time travel'' enabling chained RPCs to be executed as a single request. Additionally, Capn'Proto offers a byte-packing method that reduces payload size, albeit with the expense of some increase in serialization time. In our experiments, we refer to the byte packed version of Capn'Proto as "capnp-packed" to differentiate it from the unpacked version, "capnp".

Avro~\cite{noauthor_apache_nodate-1} is a schema-based binary serialization technology developed by Apache. Avro uses JSON to define schema data structures and namespaces. These schemas are shared between both producer and consumer. One of Avro's key advantages is its dynamic schema definition, which does not require code generation, unlike competitors such as ProtoBuf. Avro messages are also self-describing, meaning they can be decoded without needing access to the original schema. 

Additionally, we also considered the PSON format~\cite{luis_pson_2021} and Zerializer~\cite{wolnikowski_zerializer_2021}. PSON is a binary serialization format with a current implementation limited to C++ and lacks Python bindings, which restricts its applicability for our study. Zerializer, on the other hand, necessitates a specific hardware setup for implementation, placing it outside the scope of our study due to practical constraints. Consequently, while these formats might offer potential advantages, their limitations in terms of language support and hardware requirements precluded their inclusion in our experimental evaluation.

A summary of serialization protocols can be found in Table~\ref{table:protocols}. The text-based formats represent data using a text-based markup. While human-readable, text-based formats suffer from larger payload sizes and serialization costs due to the overhead of the markup describing the data. In contrast, binary formats serialize the data to bytes before transmission. These formats are not human-readable, but achieve a better payload size with lower serialization costs. Protocol-based formats also encode data in binary, but differ in that they rely on a predefined protocol definition shared between sender and receiver. Using a shared protocol frees more information out of the transmitted packet, yielding smaller payloads and faster serialization time.

\subsection{Data Streaming Technologies}
\label{subsec:streaming-protos}

\begin{table*}[t]
\centering
\caption{A comparison of different data streaming technologies.}

\setlength{\tabcolsep}{3pt}

\begin{tabular}{|l|c|c|c|c|c|c|c|c|}

    \hline
    
    \thead{Name} & 
    \thead{Type} & 
    \thead{Queue\\Mode} & 
    \thead{Consume\\Mode} & 
    \thead{Broker\\Architecture} &
    \thead{Delivery\\Guarantee} & 
    \thead{Order\\Guarantee} & 
    \thead{Code\\Generated\\Protocol} & 
    \thead{Multiple\\Consumer}  \\
    
    \hline
    
    ActiveMQ \cite{ActiveMQ} & Messaging & Pub/Sub \& P2P & Pull & Controller-Worker & at-least-once & queue-order & \ding{53} & \checkmark  \\

    Kafka \cite{ApacheKafka} & Messaging & P2P & Pull & P2P & All & partition-order & \ding{53} & \checkmark  \\
    
    Pulsar \cite{ApachePulsar} & Messaging & P2P & Push & P2P & All & global-order & \ding{53} & \checkmark  \\
    
    RabbitMQ \cite{RabbitMQ} & Messaging & Pub/Sub & Push/Pull & Controller-Worker & At-least/most-once & None & \ding{53} & \checkmark  \\
    
    RocketMQ \cite{RocketMQ} & Messaging & Pub/Sub & Push/Pull & Controller-Worker & At-least-once & queue-order & \ding{53} & \checkmark  \\
    
    Avro \cite{ApacheAvro} & RPC &  P2P & Pull & Brokerless & None & global-order & \ding{53} & \ding{53}  \\
    
    Capn'Proto \cite{CapProto} & RPC & P2P & Pull &  Brokerless & None & global-order  & \checkmark & \ding{53}  \\
    
    gRPC \cite{GRPC} & RPC & P2P & Pull &  Brokerless  & None & global-order & \checkmark & \ding{53}  \\
    
    Thrift \cite{ApacheThrift} & RPC & P2P & Pull &  Brokerless  & None & global-order & \checkmark & \ding{53}  \\
    
    ZMQ \cite{ZeroMQ} & Low Level & P2P & Pull &  Brokerless  & None & global-order & \ding{53} & \ding{53}  \\
    
    ADIOS2 \cite{godoyADIOSAdaptableInput2020} & Low Level & P2P & Pull &  Brokerless  & None & global-order & \ding{53} & \ding{53}  \\
    
    \hline
    
\end{tabular}
\label{table:messaging-technologies}
\end{table*}

In this section, we discuss three different categories of data streaming technologies: message queue-based, RPC-based, and low-level. 

\subsubsection{Message Queues}

ActiveMQ~\cite{ActiveMQ}, developed in Java by Apache, is a flexible messaging system designed to support various communication protocols, including AMQP, STOMP, REST, XMPP, and OpenWire. The system's architecture is based on a controller-worker model, where the controller broker is synchronized with worker brokers. The system operates in two modes: topic mode and queuing mode. In topic mode, ActiveMQ employs a publish-subscribe (pub/sub) mechanism, where messages are transient, and delivery is not guaranteed. Conversely, in queue mode, ActiveMQ utilizes point-to-point messaging approach, storing messages on disk or in a database to ensure at-least-once delivery. For our experiments, we utilize the STOMP communication protocol.

Kafka~\cite{ApacheKafka} is a distributed event processing platform written in Scala and Java; initially developed by LinkedIn and now maintained by Apache. Kafka leverages the concept of topics and partitions to achieve parallelism and reliability. Consumers can subscribe to one more topic, with each topic divided into multiple partitions. Each partition is read by a single consumer, ensuring message order within that partition. For enhanced reliability, topics and partitions are replicated across multiple brokers within a cluster. Kafka employs a peer-to-peer (P2P) architecture to synchronize brokers, with no single broker taking precedence over other brokers. Zookeeper~\cite{hunt_zookeeper_nodate} manages brokers within the cluster. Kafka uses TCP for communication between message queues and supports only push-based message delivery to consumers while persisting messages to disk for durability and fault tolerance.

RabbitMQ~\cite{RabbitMQ}, developed by VMWare, is a widely used messaging system known for its robust support for various messaging protocols, including AMQP, STOMP, and MQTT. Implemented in Erlang programming language, RabbitMQ leverages Erlang's inherent support for distributed computation, eliminating the need for a separate cluster manager. A RabbitMQ cluster consists of multiple brokers, each hosting an exchange and multiple queues. The exchange is bound to one queue per broker, with queues synchronized across brokers. One queue acts as the controller, while the others function as workers. RabbitMQ supports point-to-point communication and both push and pull consumer modes. Although message ordering is not guaranteed, RabbitMQ provides at-least-once and at-most-once delivery guarantees. RabbitMQ faces poor scalability issues due to the need to replicate each queue on every broker. Our experiments utilize the STOMP protocol for communication with the \textit{pika} python package. 

RocketMQ~\cite{RocketMQ}, developed by Alibaba and written in Java, is a messaging system that employs a bespoke communication protocol. It defines a set of topics, each internally split into a set of queues. Each queue is hosted on a separate broker within the cluster, and queues are replicated using a controller-worker paradigm. Brokers can dynamically register with a name server, which manages cluster and query routing. RocketMQ guarantees message ordering, and supports at-least-once delivery. Consumers may receive messages from RocketMQ either using push or pull modes. Message queuing is implemented using the pub/sub paradigm, and RocketMQ scales well with a large number of topics and consumers. 

Pulsar~\cite{ApachePulsar}, created by Yahoo and now maintained by Apache, is implemented in Java and designed to support a large number of consumers and topics while ensuring high reliability. Pulsar's innovative architecture separates message storage from the message broker. A cluster of brokers is managed by a load balancer (Zookeeper). Similar to Kafka, each topic is split into partitions. However, instead of storing messages within partitions on the broker, Pulser stores partition references in bookies. These bookies are coordinated by a bookkeeper, which is also load-balanced using Zookeeper. Each partition is further split into several segments and distributed across different bookies. The separation of message storage from message brokers means that if an individual broker fails, it can be replaced with another broker without loss of information. Similarly, if a bookie fails, the replica information stored in other bookies can take over, ensuring data integrity. Pulsar's architecture allows it to offer a global ordering and delivery guarantee, although this high reliability and scalability come at the cost of extra communication overhead between brokers and bookies.

For a detailed overview of different message queue technologies, please refer~ \cite{fu_fair_2021}. 

\subsubsection{RPC Based}
gRPC \cite{GRPC}, developed by Google, is an RPC framework that utilizes ProtoBuf as its default serialization protocol. To define the available RPC calls for a client, gRPC requires a protocol definition written in ProtoBuf. While ProtoBuf is the standard, sending arbitrary bytes from other serialization protocols over gRPC is possible by defining a message type with a bytes field. The Python gRPC implementation supports synchronous and asynchronous (asyncio) communication. For all our experiments with gRPC, we use asynchronous communication.

Capn'Proto~\cite{CapProto} and Thrift also have their own RPC frameworks. Similar to gRPC, these frameworks define remote procedural calls within their protocol definitions, using their own syntax specification. Like gRPC, they allow the transmission of arbitrary bytes by defining a message with a bytes field. 

Avro provides RPC-based communication protocol as well. Unlike other RPC-based methods, Avro does not require the RPC protocol to be explicitly defined. This flexibility comes at the expense of stricter type validation, setting Avro apart from systems such as gRPC and Thrift.

\subsubsection{Low Level}
In addition to RPC and messaging systems, we consider two low-level communication systems:  ZeroMQ and ADIOS2. Like RPC systems, they do not rely on an intermediate broker for message transmission. 

ZeroMQ (ZMQ)~\cite{ZeroMQ} is a brokerless communications library developed by iMatix. It is a highly flexible message framework that uses TCP sockets and supports various messaging patterns, such as push/pull, pub/sub, request/reply, and many more. Notably, ZeroMQ's zero-copy feature minimizes the copying of bytes during data transmission, making it well-suited for handling large messages. In our experiments, we implement a simple push/pull messaging pattern to avoid the additional communication overhead associated with RPC methods.

The ADaptable Input Output System (ADIOS)~\cite{godoyADIOSAdaptableInput2020} is a unified communications library developed as part of the U.S Department of Energy's (DoE) Exascale Computing Project. It is designed to stream exascale data loads for interprocess and wide area network (WAN) communication. In this study, we compare the WAN capabilities of ADIOS, which uses ZeroMQ for it's messaging protocol. We use ADIOS2 for communication and the low-level Python API to facilitate communication between client and server. 

We do not consider other RPC systems such as Apache Arrow Flight \cite{noauthor_arrow_nodate} which rely on ProtoBuf and gRPC for their underlying communication protocols.

A summary of the comparison of various data streaming technologies can be found in Table~\ref{table:messaging-technologies}.
Message queue-based technologies use message queues and a publish/subscribe model to transmit data. Producers publish messages to a topic, and multiple consumers can subscribe to these topics to read messages from the queue. These systems operate in \textit{push} mode, where the system delivers messages to consumers, or in \textit{pull} mode, where consumers request messages from the message queuing system. 
RPC-based technologies define a communication protocol shared between producers and consumers, eliminating the need for an intermediate broker. Producers respond to remote procedure calls (consumer requests) to provide data.
Low-level communication protocols and the ADIOS also do not require an intermediate broker. Unlike RPC technologies, they do not wait for clients' requests to send messages, reducing communication overhead. ZeroMQ and ADIOS support zero-copy messaging transfer of raw bytes, which is particularly beneficial for large array workloads where serializing and copying data can be costly.

These technologies differ in their fault tolerance. Message queuing systems prioritize reliability by caching messages to disk to prevent load shedding during high message rates. In contrast, RPC systems keep all requests in memory, offering faster performance at the expense of lower fault tolerance. Many protocol-based serialization formats introduced earlier include RPC communications libraries that support sending arbitrary bytes. For example, Protobuf-encoded messages can be sent using Avro RPC communication library.

\section{Empirical Study Design}
\label{sec:empirical-study-design}
%\subsection{Methodology}

%% draw.io: https://app.diagrams.net/#G1_Sbv8Fd2hbjwD2N8KCWTBlAJhDoyfp0m
\begin{figure*}[h]
    \centering
    \includegraphics[width=1.0\textwidth]{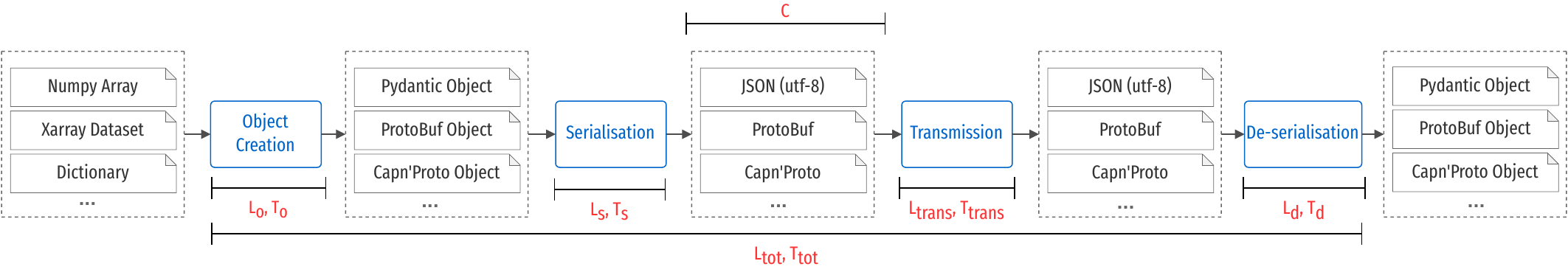}
    \caption{Illustrates the data flow from producer to consumer, indicating the places at which various performance metrics are recorded. These metrics include 
    (1) $L_o$: object creation latency, 
    (2) $T_o$: object creation throughput, 
    (3) $C$: compression ratio, 
    (4) $L_s$: serialization latency, 
    (5) $L_d$ deserialization latency, 
    (6) $T_s$: serialization throughput,
    (7) $T_d$: deserialization throughput,
    (8) $L_{trans}$: transmission latency, 
    (9) $T_{trans}$: transmission throughput, 
    (10) $L_{tot}$: total latency, and 
    (11) $T_{tot}$: total throughput.}
    \label{fig:streaming-measurements}
\end{figure*}

%% draw.io: https://app.diagrams.net/#G1qzfBbSqfFChMJf-wp80ngLmhf_L4FczH
\begin{figure*}[h]
    \centering
    \includegraphics[width=1.0\textwidth]{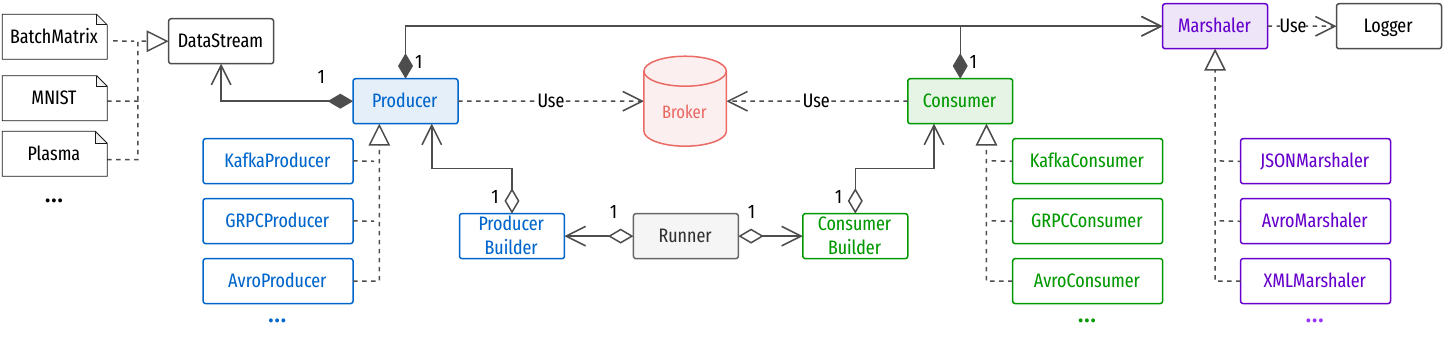}
    \caption{Diagram showing the architecture of our streaming framework. A Runner is used to create a Producer and Consumer pair for each type of streaming technology. Both producer and consumer are instantiated with a Marshaler that encodes data to the desired format (e.g. JSON, ProtoBuf, etc.). Producers are created with a data stream object that generates data samples for transmission. Depending on the streaming method, the Consumer and Producer may connect to an external message broker.}
    \label{fig:streaming-arch}
\end{figure*}

The objective of this empirical study is to investigate and compare various streaming technologies and serialization protocols for scientific data. We examine the interplay between serialization protocol and streaming technology by exploring different combinations of them. We conduct experiments on all the technologies discussed in section \ref{sec:background}, which includes 11 different streaming technologies and 13 different serialization protocols. We test each combination of technology across eight different payloads (data types), resulting in $11 \times 13 \times 8 = 1144$ different combinations. 

Finally, in we also investigate the effect of batch size on the throughput. Grouping data into batches is a common requirement during machine learning training, and we show increasing the batch size while lowering the number of communications has a positive effect on throughput.

\subsection{Performance Metrics} 

We evaluate 11 performance metrics. The first seven metrics are associated with serialization protocol, and the remaining four are linked to the combination of streaming technology and serialization protocol. 
To define these metrics, we first establish the different data sizes at various stages of the pipeline. $S_d$ denote the size of the original data, $S_o$ the size after serialization (e.g., the size after creating a gRPC object), and $S_p$ the size of the payload after serializing it to bytes. Additionally, we define the number of samples in the dataset as $N$.

The metrics we consider are:

% \begin{enumerate}[label=(\Alph*)]
\begin{enumerate} 
    \item \textbf{Object Creation Latency} ($L_o$) measures the total time taken to convert the program-specific native format (e.g. Numpy array, Xarray dataset) into the format required for transmission. This is an important metric because some formats, such as Capn'Proto store their data internally in a serialization-ready format. However, in reality, we often need to work with arrays that are in an analysis-ready format, such as Numpy array or Xarray dataset. Converting between the two models can incur a penalty since it may involve copying the data.
       
    \item \textbf{Object Creation Throughput} ($T_o = \frac{S_d}{L_o}$) 
    measures the total size of a native data type $S_d$ (e.g., Numpy array, or Xarray dataset) divided by the total time to convert it to the transmission format expected by the protocol (e.g., a ProtoBuf object or a Capn'Proto object).
    
    \item \textbf{Compression Ratio} ($C = \frac{S_p}{S_o} \times 100$) is defined as the ratio of the size of the payload $S_p$ after serialization to the size of the object $S_o$. A smaller compression ratio ultimately means less data to be transmitted over the wire, and therefore, protocols that produce a smaller payload should be more performant.
    
    \item \textbf{Serialization Latency} ($L_s$) is the total time taken to encode the original data into the serialized format for transmission. Serializing data with any protocol incurs a non-zero cost due to the need to format, copy, and compress data for transmission. A larger serialization time can potentially negate the benefits of a smaller payload size because it increases the total transmission time.
    
    \item \textbf{Deserialization Latency} ($L_d$) is similar to serialization time, this metric measures the total time required to deserialize a payload \textit{after} transmission across the wire. As with serialization time, a slow deserialization time may negate the effects of a smaller payload.

    \item \textbf{Serialization Throughput} ($T_s = \frac{S_o}{L_s}$) is the serialization time divided by the size of the object to be transmitted. This measures how many bytes per second a serialization protocol can handle, independent of the size of the data stream. 

    \item \textbf{Deserialization Throughput} ($T_d = \frac{S_o}{L_d}$) is the deserialization time divided by the size of the object received. This measures how many bytes per second a deserialization protocol can handle, independent of the size of the data stream.
\end{enumerate}

For streaming technologies, we consider two different performance metrics:

% \begin{enumerate}[label=(\Alph*), start=8]
\begin{enumerate}[, start=8]
    \item \textbf{Transmission Latency} ($L_{trans}$) is the time taken for a payload to be sent over the wire, excluding the time taken to encode the message.
    
    \item \textbf{Transmission Throughput} ($T_{trans} = \frac{S_d}{L_{trans}}$) is similar to total throughput, but considers the payload size divided by the time taken to send the message over the wire, exclusive of the serialization time.

    \item \textbf{Total Latency} ($L_{tot}$) is the total time for a payload to be transmitted from producer to consumer, inclusive of the serialization time.
    
    \item \textbf{Total Throughput} ($T_{tot} = \frac{S_d}{L_{tot}}$) is the original data object size divided by the total time to send the message. Throughput measures the rate of bytes that can be communicated over the wire.

\end{enumerate}

We make a distinction between \textit{transmission time} and \textit{total time} (Figure~\ref{fig:streaming-measurements}). The total time is the end-to-end transmission of a message, including the time to serialize the message and send it over the wire. Transmission time is the time taken to transmit the payload \textit{excluding} the serialization and deserialization times. Similarly, we can calculate total and transmission throughput.

\subsection{Dataset}
\label{subsubsec:payloads}

In our experiments, we consider eight different payloads, ranging from simple data to common machine learning workloads, and includes data from the fusion science domain. Our goal is to cover a range of scenarios. This section briefly describes the datasets used to evaluate performance with various streaming technologies and serialization protocols.

\begin{enumerate}
    \item[1--3)] \textbf{Numerical Primitives} refer to basic data types that represent fundamental numerical values, e.g., NumPy data types\cite{numpy-data-types}. As a baseline comparison, we use randomly generated numerical primitives for (1) int32, (2) float32, and (3) float64.
    
    \addtocounter{enumi}{3} % increment the counter by 2 

    \item \textbf{BatchMatrix} is a synthetic dataset where each message consists of a randomly generated 3D tensor of type float32 with shape $\{32, 100, 100\}$ to simulate sending a batched set of image samples. The random nature of the synthetic data makes it incompressible.

    \item \textbf{Iris Data} is a dataset using the well-known Iris dataset~\cite{fisherUseMultipleMeasurements1936}. The Iris dataset contains an array of four float32 features and a one-dimensional string target variable.
    
    \item \textbf{MNIST} is the widely used machine learning image dataset~\cite{deng2012mnist} as a realistic example of streaming 2D tensor data.
    
    \item \textbf{Scientific Papers} dataset is a well-known dataset in the field of NLP and text processing~\cite{Cohan_2018}. The dataset comprises 349,128 articles of text from PubMed and arXiv publications. Each sample is repeated as a collection of string for properties such as article, abstract, and section names for transmission.
    
    \item \textbf{Plasma Current Data} is a realistic example of scientific data, we use plasma current data from the MAST tokamak~\cite{sykesFirstPhysicsResults2001}. Each item of plasma current data contains three 1D arrays of type float32: data, time, and errors. The data array represents the amount of current at each timestep, the time represents the time the measurement was taken in seconds, and the errorsrepresent the standard deviation of the error in the measured current.
\end{enumerate}

\subsection{Implementation and Experimental Setup}
\label{subsec:implementation}

We developed a framework to measure the performance of streaming and serialization technology. The architecture diagram of our framework is shown in Figure~\ref{fig:streaming-arch}, which follows service-oriented architecture~\cite{Khan:DIM:2022, Khan:TSC:2022} and is implemented in Python. We used the appropriate Python client library for each streaming and serialization technology. The source code can be found in our GitHub repository~\cite{github-streaming-perf}.

The user interacts with the framework through a command-line interface. A test runner sets up both the server-side and client-side of the streaming test. 

The server side requires the configuration of three components:

\begin{itemize}
    \item \textbf{DataStream} component handles loading data for transmission. This can be any one of the payloads described in section~\ref{subsubsec:payloads}.
    
    \item \textbf{Producer} functions as the server side of the application. It packages data from the selected data stream and transmits it over the wire using the selected streaming technology, which may be any of the technologies described in section~\ref{subsec:streaming-protos}.
    
    \item \textbf{Marshaler} handles the serialization of the data from the stream using the specified serialization protocol. This can be any of those described in section~\ref{subsec:serialization-protos}.
\end{itemize}

The configuration of the client side is similar but only requires a marshaler to be configured to match the one used for the producer. It does not require knowledge of the data stream. 

\begin{itemize}
    \item \textbf{Consumer} functions as the client side of the application. It receives data transmitted by the producer using the selected streaming technology, processes the incoming messages, and performs the necessary actions.
    Producers and consumers interact using a configured protocol.

    \item \textbf{Broker} required by the streaming protocol (e.g., for Kafka, RabbitMQ, etc.) are run externally from the test in the background. In our framework, we configure all brokers using docker-compose \cite{merkel2014docker} to ensure that our broker configurations are reproducible for every test.

    \item \textbf{Logger} is used by the marshaler to capture performance metrics for each test in a JSON file. For each message sent, the logger captures four timestamps: 1) before serialization, 2) after serialization, 3) after transmission, and 4) after deserialization. Using these four timestamps, we can calculate the serialization, deserialization, transmission, and total time. Additionally, the logger captures the payload size of each message immediately after serialization. With this additional information, we can calculate the average payload size and throughput of the streaming service.

\end{itemize} 

 ADIOS and ZeroMQ can directly send array data without copying the input array. However, to achieve this, the array data must be directly passed to the communication library without serialization. Therefore, we additionally consider ZeroMQ and ADIOS to have their own ``native'' serializtion strategy for each stream, which is only used with their respective streaming protocol. This allows for a fair comparison with other technologies because sending a serialized array with ADIOS or ZeroMQ incurs an additional copy that could be circumvented by properly using their native zero-copy functionality.

Following the convention of previous work \cite{fu_fair_2021, khanIEEEAccess2023}, we run each streaming test locally, with the producer and consumer on the same machine to avoid network-specific issues. 

\section{Results}
\label{sec:results}

In this section we present the results of our experiments with the combination of different streaming technologies, serialization protocols, and data types. In this study, we used datasets derived from various data analysis frameworks, e.g., NumPy, Xarray, etc.

\hfill \break
\noindent \textbf{1) Object Creation Latency --}

Some serialization protocols, such as Cap'n Proto and Protocol Buffers (Protobuf), require data to be converted from its native format. This data conversion introduces a performance overhead. Conversely, serialization protocols such as JSON, BSON, and Pickle, which do not require format alterations, allow for direct storage of data within a Pydantic class structure. The former approach minimizes data manipulation and potentially reduces processing time.

\begin{figure*}[]
    \centering
    \includegraphics[width=1.\textwidth]{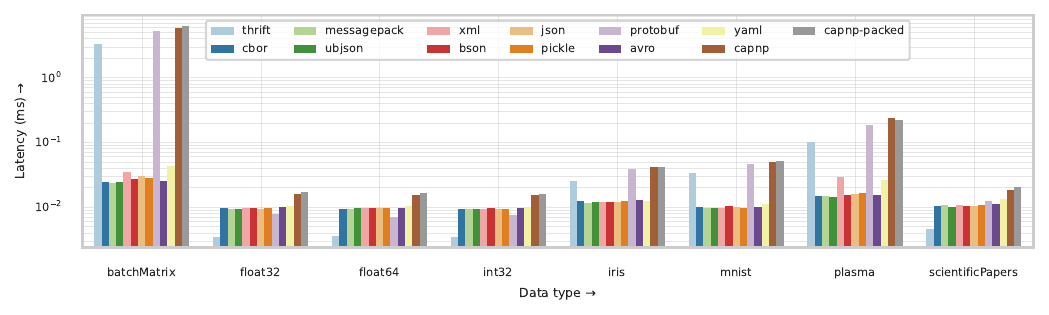}
    \caption{Object creation latency ($L_o$), measured in milliseconds (ms), of various data types arranged in the x-axis and serialization methods shown in colored bars}.
    \label{fig:object-creation-latency}
\end{figure*}

Figure~\ref{fig:object-creation-latency} shows object creation latency across different serialization methods. The results demonstrate that protocols such as Protobuf, Thrift, and Cap'n Proto exhibit higher object creation latencies for larger array datasets like BatchMatrix, plasma, and MNIST. This increased latency is attributed to the necessary data copying process, where data must be transferred into the protocol-specific data types.

\hfill \break
\noindent \textbf{2) Object Creation Throughput --}

Object creation throughput measures the time required to convert data from its native structure (such as a NumPy array) into a serialization format, normalized by the size of the data. This metric is important when the format used for processing the data differs from the format used for transmission. In such cases, object creation often necessitates copying the data, which can significantly affect overall throughput, particularly for large datasets.

\begin{figure*}[]
    \centering
    \includegraphics[width=1.\textwidth]{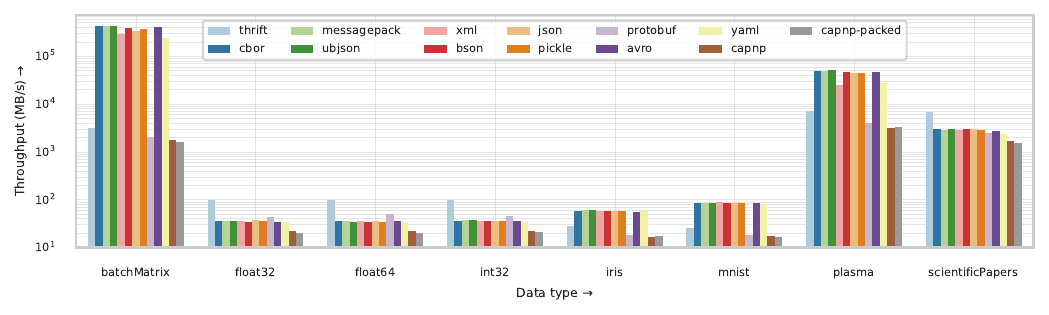}
    \caption{Object creation throughput, ($T_o$) measured in megabytes per second (MB/s), of various data types arranged in the x-axis and serialization methods shown in colored bars}.
    \label{fig:object-creation-throughput}
\end{figure*}

Figure~\ref{fig:object-creation-throughput} shows the object creation throughput for each dataset and each serialization method. Protocol based methods incur a higher penalty for the object creation. This penalty is more noticeable in larger datasets such as the BatchMatrix plasma datasets, and scientific papers.

\begin{figure*}[]
    \centering
    \includegraphics{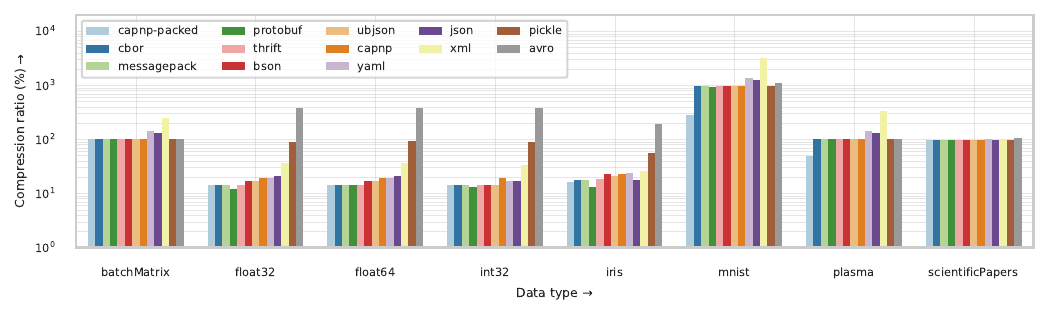}
    \caption{The compression ratio ($C$) of various data types arranged in the x-axis and serialization methods shown in colored bars}.
    \label{fig:compression-ratio}
\end{figure*}

\hfill \break
\noindent \textbf{3) Compression Ratio --}
This performance metric quantifies the efficiency of serialization by measuring the ratio between the original data size and the size of the serialized payload. This metric remains unaffected by the choice of streaming protocols.

Figure~\ref{fig:compression-ratio} shows the compression ratio of serialization protocols. Pickle, Avro, and XML consistently produce largest serialized payloads, often exceeding the original data size. This inefficiency is due to their text-based serialization and the additional metadata tags that contribute to overhead. Pickle, despite being a binary format designed for storing Python objects, is particularly inefficient in terms of size, making it suboptimal for data streaming.

The result show that binary compression algorithms perform best amongst all options. capnp-packed, Protobuf, CBOR, BSON, UBJSON, Thrift are all competitive in terms of compression ratio. The reason behind this performance can be attributed to their ability to achieve near-identical compression, which is close to the limits of what is possible for that particular data stream.

Examining across data types, we can see that the BatchMatrix dataset is fundamentally limited. This is because it is made up of randomly generated numbers, making it incompressible due to the lack of redundancy in the data. For more realistic data such as MNIST and plasma, more variety in compression ratio is achieved. Data redundancy can be exploited to achieve a better compression ratio. For example capnp-packed. Text-based formats, such as YAML, JSON, XML, and Avro, achieve significantly worse compression ratio. In fact, due to the extra markup required for these formats, they can produce a larger payload size that the original data.

\begin{figure*}[]
    \centering
    \includegraphics[width=1\textwidth]{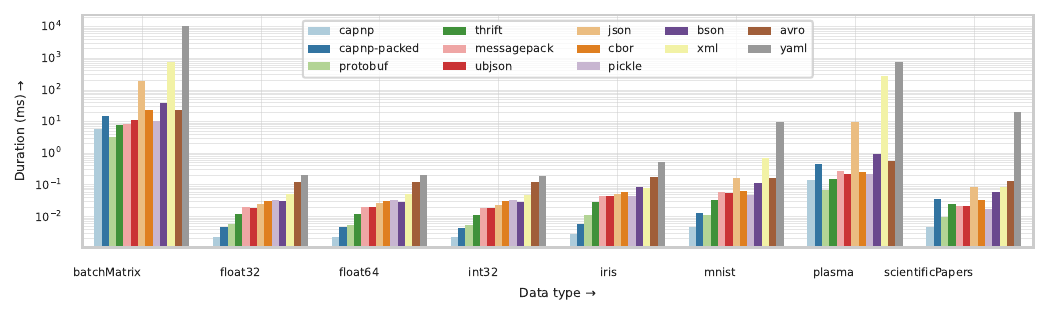}
    \caption{Serialization ($L_s$) latency of various data types arranged in the x-axis and serialization methods shown in colored bars. Protocol serialization methods such as Protobuf and Captn'Proto consistently offer the best performance in terms of both serialization. Text-based serializers (YAML, XML, etc.) add a large latency penalty to serialization by increasing the verbosity of the data.}
    \label{fig:serialisation-time}
\end{figure*}

\begin{figure*}[]
    \centering
    \includegraphics[width=1\textwidth]{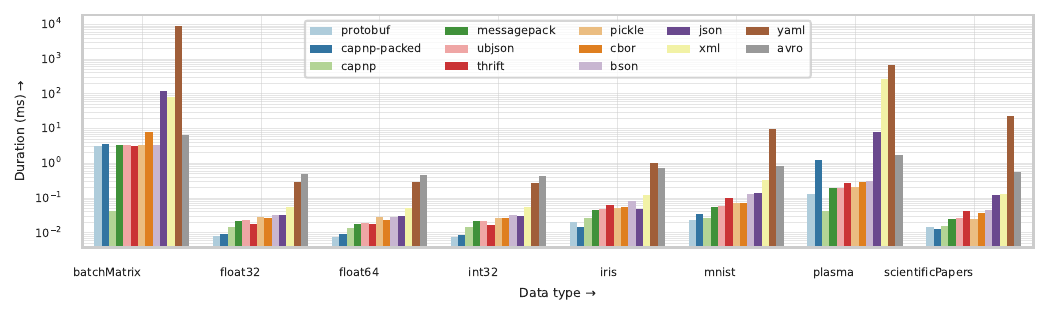}
    \caption{Deserialization ($L_d$) latency of various data types arranged in the x-axis and serialization methods shown in colored bars. As with serialization, protocol methods such as Protobuf and Captn'Proto offer the best performance. Likewise, Text-based serialization methods (YAML, XML, etc.) add a large latency penalty by increasing the verbosity of the data.}
    \label{fig:deserialisation-time}
\end{figure*}

\hfill \break
\noindent \textbf{4) Serialization Latency --}

The results for serialization time are shown in Figure~\ref{fig:serialisation-time}. We observed a clear trend across all data types, with text-based protocols (e.g., Avro, YAML, etc.) showing the slowest serialization time, while binary-encoded protocol-based methods (e.g., Capn'Proto, Protobuf, etc.) demonstrate the fastest. Binary-encoded methods without protocol support methods fall between these two extremes.

Capn'Proto achieves the fastest serialization times among all methods. This performance advantage is likely due to Capn'Proto design, which stores data in a format that is ready for serialization.

\hfill \break
\noindent \textbf{5) Deserialization Latency --}

Figure \ref{fig:deserialisation-time} shows the results for deserialization latency. We see a clear trend across all data types, with text-based protocols displaying longer deserialization times compared to binary-encoded, protocol-based method.

Like serialization, Capn'Proto consistently achieves the fastest deserialization times across all tests. This superior performance is likely due to its design, which stores data in a format that is already optimized for serialization and deserialization, reducing the need for additional processing.

\begin{figure}[!htb]
    \centering
    \includegraphics[width=1.0\columnwidth]{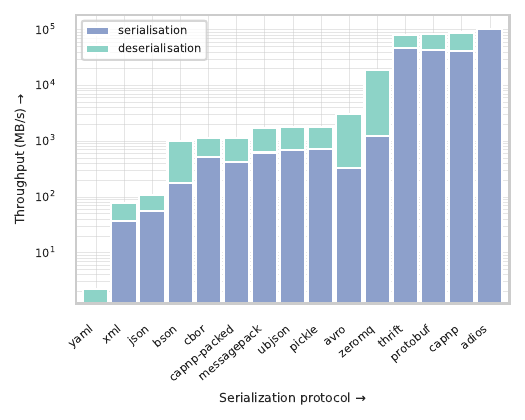}
    \caption{Serialization ($T_s$) and deserialization ($T_d$) throughput for each serializer averaged over all data types.}
    \label{fig:serialisation-throughput}
\end{figure}

\hfill \break
\noindent \textbf{6) Serialization Throughput --}
Figure~\ref{fig:serialisation-throughput} shows the average throughput for serialization of the data using various protocols. The results show that protocols-based (e.g., ProtoBuf, Thrift, and Capn'Proto) serialization techniques achieve the highest throughput. Binary methods that are protocol-independent achieve moderate throughput. Text-based methods perform the worst due to their high serialization overhead.

Surprisingly, Avro also performs well, although it is human-readable text-based method. Its protocol-based nature likely contributes to this efficiency - this implies that both the producer and consumer are aware of the types and structures being transmitted, facilitating faster throughput.

\hfill \break
\noindent \textbf{7) Deserialization Throughput --}
Figure~\ref{fig:serialisation-throughput} also shows the average throughput for deserialization of the data using various protocols. The results show that deserialization throughput is consistently lower across all methods, suggesting that deserialization is a significant bottleneck in data transmission process.

\begin{figure*}[]
    \centering
    \includegraphics[width=1.\textwidth]{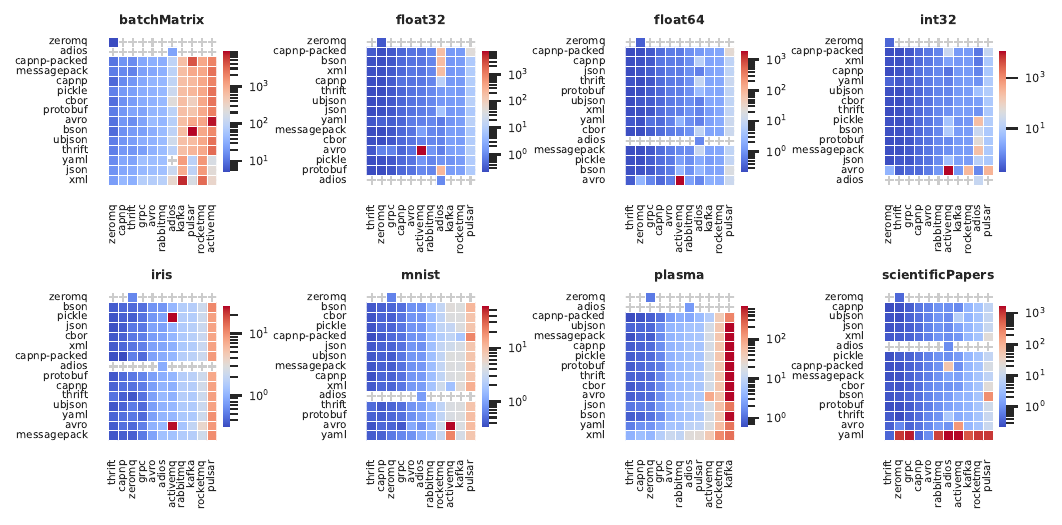}
    \caption{
    Each heatmap shows transmission latency ($L_{trans}$) for each combination of serialization protocols (vertical axis) and streaming technologies (horizontal axis). Dark red indicates higher latency. The left-to-right trend (changes in color of the heatmap) indicates that the choice of streaming technology has the most significant impact on latency.
    }
    \label{fig:transmission-latency.pdf}
\end{figure*}

\hfill \break
\noindent \textbf{8) Transmission Latency --}
Figure~\ref{fig:transmission-latency.pdf} shows the transmission latency for various combinations of serialization and streaming technologies. In the heatmap, combinations are sorted by the average latency from lowest to highest for each streaming technology.

The results show that the transmission latency mainly depends on the choice of streaming technology rather than serialization protocol.Streaming technologies, which require a broker as an intermediary, introduce latency. In contrast, RPC-based technologies, which operate without a broker, achieve lower latency. Among messaging technologies, RabbitMQ performs better with larger payloads, while ActiveMQ achieves lower latency for smaller payloads but struggles with largest payloads (e.g., BatchMatrix). In RPC-based methods, Thrift consistently delivers the lowest latency except for the BatchMatrix stream, where Capn'Proto narrowly beats Thrift.

For larger payloads such as BatchMatrix and Plasma data types, the choice of serialization protocol becomes more noticeable. While it is difficult to identify a trend in latency across serialization protocols, it is clear that XML and YAML are inefficient for handling larger payloads.

For the BatchMatrix data, an issue arises when attempting to send a large YAML-encoded payload through the Python API, which causes a segmentation fault in ADIOS. As a result, subsequent latency and throughput measurements result in NaN values, which are represented as empty cells in Figure~\ref{fig:transmission-latency.pdf}.

\begin{figure*}[]
    \centering
    \includegraphics[width=1.\textwidth]{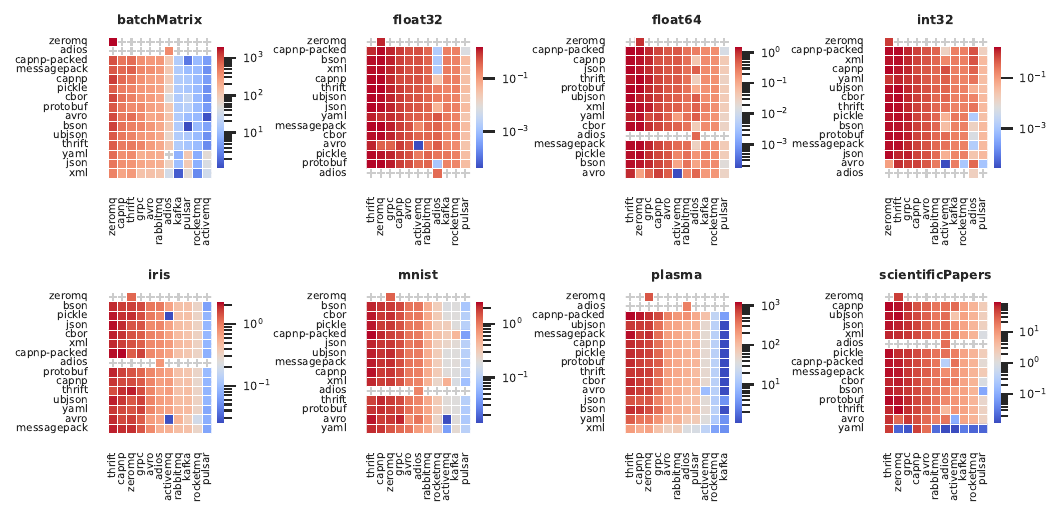}
    \caption{Each heatmap shows transmission throughput ($T_{trans}$) for different serialization protocols (vertical axis) and streaming technologies (horizontal axis). The left-to-right trend (changes in color of the heatmap) indicates that the choice of streaming technology has the most impact than serialization protocol.}
    \label{fig:transmission-throughput}
\end{figure*}

\hfill \break
\noindent \textbf{9) Transmission Throughput --}
Figure~\ref{fig:transmission-throughput} shows that RPC methods achieve higher transmission throughput. When handling larger payloads, such as the BatchMatrix and plasma data, protocol-based serialization methods such as Thirft, Capn'Proto, and Protobuf deliver higher throughput. Interestingly, MessagePack also performs well with larger payloads.

Similar to latency, the choice of streaming technology plays an important role than the choice of serialization method. However, a trend favouring protocol-based serialization methods emerges with some larger datasets, such as the plasma dataset.

\begin{figure*}[]
    \centering
    \includegraphics[width=1\textwidth]{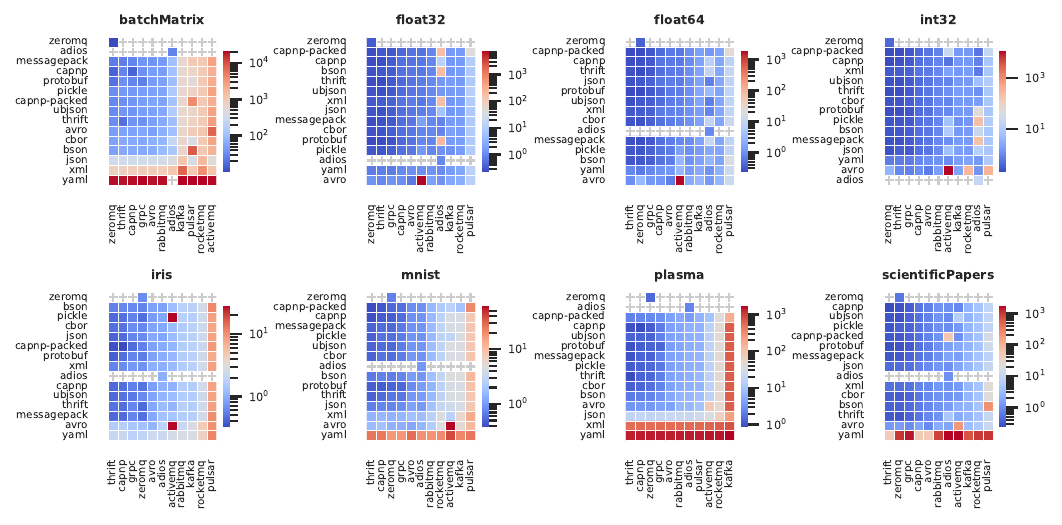}
    \caption{Each heatmap shows total latency ($L_{tot}$) for different serialization protocols (vertical axis), and streaming technologies (horizontal axis). The color variations in the heatmap, observed from top to bottom and left to right, indicate that both streaming technology and serialization protocol affect overall system performance; however, streaming technology has the most significant influence.
    }
    \label{fig:total-latency}
\end{figure*}

\hfill \break
\noindent \textbf{10) Total Latency --}
Figure~\ref{fig:total-latency} shows the total latency across various methods. As observed earlier, Thrift, Capn'Proto, and ZeroMQ perform well in this metric. ZeroMQ achieves the lowest latency in the BatchMatrix data as it avoids the overhead associated with copying data into a new structure, but is necessary with Thift or Protobuf. Among the broker-based methods, RabbitMQ consistently performs well.

In terms of serialization methods, protocol-based methods generally perform the best across all datasets and streaming technologies. However, it is unclear which method offers the lowest latency overall. Protocol-based methods can achieve high throughput by integrating different serialization protocols with RPC frameworks. For example, in case of the MNIST dataset, Capn'Proto achieves the lowest latency when used with the thrift protocol.

A clear trend emerges in favor of protocol serialization for complex datasets such as Iris, MNIST, and plasma.  Among streaming technologies, Thrift generally shows the best performance.

\begin{figure*}[]
    \centering
    \includegraphics[width=1.\textwidth]{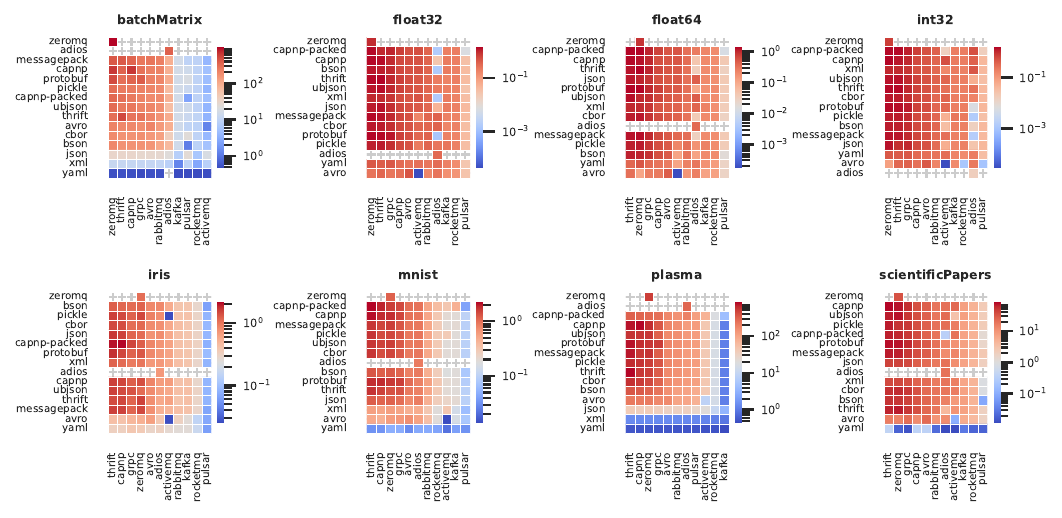}
    \caption{Each heatmap shows total throughput ($T_{tot}$) for different serialization protocols (vertical axis) and streaming technologies (horizontal axis).  While both factors influence total throughput, streaming technology has a more significant impact.}
    \label{fig:total-throughput}
\end{figure*}

\hfill \break
\noindent \textbf{11) Total Throughput --}
Figure~\ref{fig:total-throughput} shows total throughput, which is consistent with the total latency results discussed earlier. Protocol-based methods achieve the highest throughput, with Thrift emerging as the best-performing serialization protocol. ZeroMQ performs particularly well with the large dataset, e.g., BatchMatrix and plasma.
While no single serialization protocol conclusively outperforms the others, there is a trend favoring protocol-based serialization protocols, which consistently deliver the highest throughput. 

The choice of streaming technology has a more significant impact that the choice of serialization on throughput. This is largely due to the difference between broker based, which have message delivery guarantees, and RPC based systems which do not.

\begin{figure*}[]
    \centering
    \includegraphics[width=1.0\textwidth ,trim=0.0cm 0.0cm 0.0cm 0.0cm, clip]{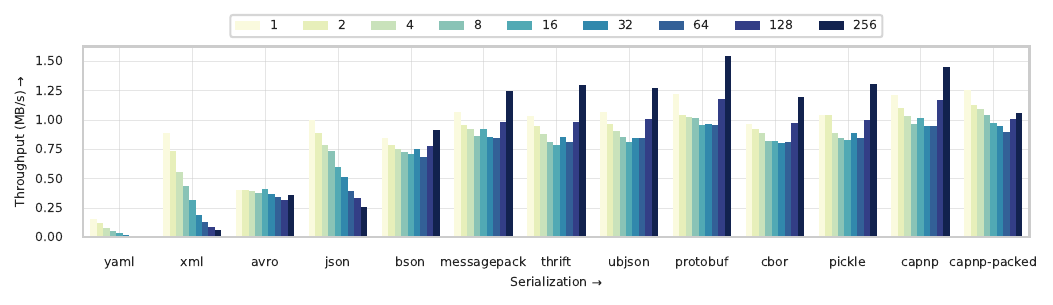}
    \caption{Total throughput ($T_{tot}$) for various batch sizes (ranging from 1 to 256) for the MNIST dataset.}
    \label{fig:batch-throughput}
\end{figure*}

\hfill \break
\noindent \textbf{12) Effect of Batch Size on Throughput --}
In machine learning applications, data is often processed in batches. 

We selected the MNIST dataset as an example for this test. Figure~\ref{fig:batch-throughput} shows the throughput of the MNIST dataset with a variable batch size. 

In most cases, when the batch size is increased beyond 32 images per batch, the overall throughput begins to improve because fewer packets are needed to be communicated over the network. 

At larger batch sizes ($>128$), the throughput continues to increase because fewer transmission time is significantly slower than the serialization/deserialization cost. So grouping many examples into a single transmission improves throughput.

Text-based protocols experience reduced throughput with increasing batch size due to their verbose nature, increased parsing complexity, and higher input-output demands. Conversely, binary and protocol-encoded formats benefit from increased batch size because they are optimized for compactness, machine efficiency, and effective use of network bandwidth. This contrast arises from the fundamental design trade-offs between human readability and machine efficiency.
This observation is consistent with previous results; generally, protocol-based methods offer the best throughput.
\section{Discussion}
\label{sec:discussion}

\subsection{Recommendations}

We found that the choice of messaging technology has a greater impact on the performance than the serialization protocol. 
RPC systems outperform messaging broker systems in terms of speed, primarily due to the overhead introduced by the intermediary broker in messaging systems, which adds extra processing and communication steps. RPC systems are more efficient for high throughput and low latency transmission of large datasets. However, they lack the robust delivery guarantees provided by messaging broker systems. 
Apache Thrift achieves high throughput and low latency across various scenarios. Among broker-based systems, RabbitMQ generally demonstrates the best performance.

Protocol-based serialization methods, such as Capt'n Proto and ProtoBuf, deliver the best performance for compressible, complex datasets, while MessagePack is a competitive choice for smaller or more random data.
Protocol-based serialization methods offer the fastest serialization and best compression, with Thrift offering the best throughput and Capn'Proto offering the best compression. Binary serialization methods offer more flexibility at the cost of slower serialization speeds. Among these, MessagePack generally performed the best. In terms of text-based protocols, JSON demonstrated the best performance due to its lightweight markup and smaller payload size compared toYAML or Avro. 
    
We observed minimal differences when combining various protocol-based serialization and messaging systems. Although we hypothesized that ProtoBuf would be most efficient when combined with the gRPC, or that Capn'Proto's RPC implementation would perform best with Capn'Proto, our results did not support that assumption.
    
For array datasets, larger batch sizes resulted in higher throughput when using either a binary or protocol-based serialization methods (Figure~\ref{fig:batch-throughput}). In contrast, text-based serialization methods, due to the additional markup and lack of compression, showed no such benefit from batching.  

\subsection{Limitations and future directions}

A key limitation of this study is that we did not investigate the potential of scaling with multiple clients. Previous research has examined this aspect in the context of message queuing systems~\cite{fu_fair_2021}. Future work could focus on examining the reliability of various RPC technologies as the number of consumers increases.  
\section{Conclusion}
\label{sec:conclusion}

In this work, we investigated 143 combinations of different serialization methods and messaging technologies, assessing their performance across 11 different metrics. Each combination was benchmarked using eight different datasets, ranging from toy datasets to machine learning data and scientific data from the fusion energy domain. We found that messaging technology has the most significant impact on performance, irrespective of the serialization method used. Protocol-based methods consistently deliver the highest throughput and lowest latency, though this comes at the cost of flexibility and robustness. 
We observed minimal differences when combining various protocol-based serialization methods and messaging systems. Lastly, we found that batch size affects the data throughput for all binary and protocol-based serialization methods.   
\section*{Contribution}

SJ: Designed and implemented the experimental framework, shaping the research methodology and contributions to the writing and conceptualization of the paper.
NC: Provided the MAST data for the study and offered expertise in the fusion domain, enhancing the scientific rigor of this empirical study and editing and refining the manuscript.
SK: Provided technical supervision - introduced the core idea and contributed to the writing and editing of the paper, figures, and plots.

\section*{Acknowledgment}

We would like to thank our colleagues at UKAEA and STFC for supporting the FAIR-MAST project. Additionally we would like to thank Stephen Dixon, Jonathan Hollocombe, Adam Parker, Lucy Kogan, and Jimmy Measures from UKAEA for assisting our understanding of the Fusion data. We would also like to extend our thanks to the wider FAIR-MAST project which include Shaun De Witt, James Hodson, Stanislas Pamela, Rob Akers from UKAEA and Jeyan Thiyagalingam from STFC. We also want to extend our gratitude to the MAST Team for their efforts in collecting and curating the raw diagnostic source data during the operation of the MAST experiment.

\bibliographystyle{IEEEtran}
\bibliography{IEEEabrv,references/references.bib}

\end{document}